\documentclass[aps,prd,twocolumn,
eqsecnum,showpacs,amsmath,nofootinbib]{revtex4}

\usepackage[dvips]{color,graphicx}
\usepackage{amsfonts,amssymb,theorem,mathrsfs,times}
\newcommand{\ma}[1]{\mbox{$\mathcal{#1}$}}
\newcommand{\mas}[1]{\mbox{$\mathscr{#1}$}}
\newcommand{\qed}{\hbox{\rule[-2pt]{6pt}{6pt}}}
\newcommand{\D}{{\rm d}}

{\theorembodyfont{\upshape}
\newtheorem{Prop}{Proposition}}
{\theorembodyfont{\upshape}
}
{\theorembodyfont{\upshape}
\newtheorem{lm}{Lemma}}
{\theorembodyfont{\upshape}
\newtheorem{dn}{Definition}}
{\theorembodyfont{\upshape}
}
{\theorembodyfont{\upshape}
}

\newcommand{\dalm}{\kern1pt\vbox{\hrule height 0.9pt\hbox{\vrule width
0.9pt\hskip 2.5pt\vbox{\vskip 5.5pt}\hskip 3pt\vrule width 0.3pt}\hrule height
0.3pt}\kern1pt}

\begin{document}

\title{
Dynamical black holes with symmetry in Einstein-Gauss-Bonnet gravity
}

\author{Masato Nozawa$^1$}
\email{nozawa@gravity.phys.waseda.ac.jp}
\author{Hideki Maeda$^{2,3}$}
\email{hideki@cecs.cl}


\address{ 
$^{1}$Department of Physics, 
Waseda University, Tokyo 169-8555, Japan,\\
$^{2}$Centro de Estudios Cient\'{\i}ficos (CECS), Arturo Prat 514, Valdivia, Chile\\
$^{3}$Department of Physics, International Christian University, 
3-10-2 Osawa, Mitaka-shi, Tokyo 181-8585, Japan
}

\date{\today}

\begin{abstract} 
We explore various aspects of dynamical black holes defined 
by a future outer trapping horizon in $n(\ge 5)$-dimensional Einstein-Gauss-Bonnet gravity.
In the present paper, we assume that the spacetime has symmetries
 corresponding to the isometries of an $(n-2)$-dimensional maximally
 symmetric space and the Gauss-Bonnet coupling constant is non-negative. 
Depending on the existence or absence of the general relativistic limit, 
solutions are classified into GR and non-GR branches, respectively. 
Assuming the null energy condition on matter fields, 
we show that a future outer trapping horizon in the GR branch 
possesses the same properties as that in general relativity.
In contrast, that in the non-GR branch is shown to be non-spacelike 
with its area non-increasing into the future.
We can recognize this peculiar behavior to arise from a fact that the null energy
 condition necessarily leads to the null convergence condition 
for radial null vectors in the GR branch, but not in the non-GR branch.
The energy balance law yields the first law of a trapping horizon, 
from which we can read off the entropy of a trapping horizon reproducing Iyer-Wald's expression.
The entropy of a future outer trapping horizon is shown to be non-decreasing in both branches along its generator. 
\end{abstract}


\maketitle


\section{Introduction}

Black holes in our universe are commonly considered to be formed by
the collapse of massive stars in the final stage of their lives. 
Since one would expect the gravitational collapse to settle down 
to equilibrium states at late times, 
stationary black holes are thus of great 
physical interest and intensively studied in the literature. 
One major achievement of the study of stationary black holes
is the uniqueness theorem: the only stationary vacuum black hole 
in an asymptotically flat spacetime is the Kerr black hole, and 
it is completely specified by its mass and angular momentum
(see e.g., \cite{carter}). 
Toward the proof of the uniqueness theorem of black holes,
the ``rigidity theorem'' 
\cite{arealaw,Hawking:1973uf,Chrusciel:1996bj,Hollands:2006rj} 
plays an essential role. 
This theorem insists that a rotating black 
hole must be axisymmetric:
the stationary black-hole event horizon is the Killing horizon. 
Killing horizons exhibit thermodynamical properties 
\cite{Bardeen1973,Hawking:1974sw,wald2,Wald:1993nt,iyerwald1994,Iyer:1995kg}, 
which strongly suggest
the intimate association between classical general relativity, 
quantum theory and statistical mechanics.

However, black holes in our universe rarely reach equilibrium.
They evolve by absorbing stars and galactic remnants, or by 
coalescing. In these fully dynamic processes, one cannot
identify the location of the event horizon at each time, because
it is determined by the global structure of spacetime.
Over the past decades, some local definitions of
horizons have provided useful and powerful implements for the
analysis of dynamical aspects of non-stationary black holes.
Among other things, {\it trapping horizons}
\cite{hayward1994,Hayward:1994yy,Hayward:2004dv,Hayward:2004fz}
defined by Hayward, 
{\it isolated horizons} \cite{Ashtekar:1999yj,Ashtekar:2000sz} 
and {\it dynamical horizons} 
\cite{Ashtekar:2002ag,Ashtekar:2003hk,Ashtekar:2004cn,Ashtekar:2005ez}
defined and developed by Ashtekhar and his coworkers, provide a quasi-local 
characterization of black holes. 
(See~\cite{booth2005} for a review of the quasi-local horizons.)
In contrast to the event horizon, neither of the above requires
the knowledge of the entire future.
Related works have revealed that 
these horizons also exhibit laws of black-hole dynamics
analogous to those of the Killing horizon, irrespective of
the highly dynamical settings. These new concepts of horizons
involve applications to numerical relativity, quantum gravity
and so on.

Gravitation physics in higher dimensions
is a prevalent subject of current research motivated by string theory. 
Higher-dimensional general relativity is obtained by the
lowest order of the Regge slope expansion of strings.
Even in general relativity, black holes in higher dimensions
expose a sharp difference from those in four dimensions \cite{hdbh}.
The next stringy compensation yields
the quadratic Riemann curvature terms in the heterotic string case \cite{Gross}. 
In order for the graviton amplitude to
be ghost-free, a special combination of the remaining curvature-squared 
terms is required to be the renormalizable Gauss-Bonnet term \cite{Zwiebach:1985uq}. 
These higher-curvature terms come into play in extremely curved 
regions. Black holes and singularities are one of the best testbeds 
for demonstrating the effects of higher curvature terms. 
To elucidate the nature of black holes and singularities in
Einstein-Gauss-Bonnet gravity will aid in understanding the 
higher-dimensional, stringy corrected theory of gravity. 
This is the main subject of the present paper.

We explore the dynamics of black holes 
in $n(\ge 5)$-dimensional Einstein-Gauss-Bonnet gravity
by taking particular notice of the trapping horizon.
(As another approach, black-hole dynamics in the framework of 
the isolated horizon were addressed in \cite{lb2007}.)
The spacetime is supposed to have symmetries corresponding to the isometries
of an $(n-2)$-dimensional maximally symmetric space, which is also assumed to
be compact to make physical quantities finite. 
The energy-momentum tensor of matter fields is left arbitrary 
except for suitable energy conditions.
Since the trapping property is inherently a local notion, it is suitable to 
manipulate basic equations by means of (quasi-)local quantities.
As shown in \cite{mn2007},
our quasi-local mass defined geometrically \cite{maeda2006b} 
makes the field equations rather tractable. This is the generalization of
the Misner-Sharp quasi-local mass \cite{ms1964} and shares 
similar properties with the four-dimensional counterpart \cite{hayward1996,mn2007}.
The mass of a trapping horizon is shown to obey 
an isoperimetric inequality similar to that of Penrose
and gives an upper or lower bound in some cases.  
Solutions in Einstein-Gauss-Bonnet 
gravity are classified into two classes in general: the GR branch (having a general 
relativistic limit) and the non-GR branch (having no general relativistic
limit). 
Our main arguments show that, under the null energy condition, 
a future outer trapping horizon in the GR branch possesses 
the same properties as that in general relativity. 
On the other hand, the non-GR-branch solutions behave rather
pathologically under the null energy condition. 
We also unveil the laws of black-hole dynamics and discuss the area and entropy laws.

The rest of the present paper is constituted as follows.
In the following section, a concise overview of Einstein-Gauss-Bonnet gravity, 
the definition of our quasi-local mass, and basic equations are given. 
Section~III focuses on the clarification of the dynamical 
properties of trapping horizons.
Various types of trapping horizons are 
scrutinized for each branch, and subsequently the black-hole dynamics are discussed. 
Concluding remarks and discussions including future prospects 
are summarized in section~IV.

Our basic notations follow \cite{wald}.
The conventions of curvature tensors are 
$[\nabla _\rho ,\nabla_\sigma]V^\mu ={R^\mu }_{\nu\rho\sigma}V^\nu$ 
and $R_{\mu \nu }={R^\rho }_{\mu \rho \nu }$.
The Minkowski metric is taken to be the mostly plus sign, and 
Roman indices run over all spacetime indices.
We adopt the units in which only the $n$-dimensional gravitational
constant $G_n$ is retained.


\section{Preliminaries}
We begin by a brief description of Einstein-Gauss-Bonnet gravity 
in the presence of a cosmological constant.
The action in $n (\geq 5)$-dimensional spacetime is given by
\begin{align}
\label{action}
S=\int\D ^nx\sqrt{-g}\biggl[\frac{1}{2\kappa_n^2}
(R-2\Lambda+\alpha{L}_{\rm GB}) \biggr]+S_{\rm matter},
\end{align}
where
$R$ and $\Lambda$ are the $n$-dimensional Ricci scalar 
and the cosmological constant, respectively. 
$S_{\rm matter}$ in Eq.~(\ref{action}) is the action for matter
fields 
and $\kappa_n := \sqrt{8\pi G_n}$, where $G_n$ is 
the $n$-dimensional gravitational constant.
The Gauss-Bonnet term $L_{\rm GB}$ comprises 
the combination of the Ricci scalar, 
Ricci tensor $R_{\mu\nu}$ and Riemann tensor ${R^\mu}_{\nu\rho\sigma}$
as
\begin{align}
{L}_{\rm GB} := R^2-4R_{\mu\nu}R^{\mu\nu}
+R_{\mu\nu\rho\sigma}R^{\mu\nu\rho\sigma}.
\end{align}
In four-dimensional spacetime, the Gauss-Bonnet term 
does not contribute to the field equations since it
becomes a total derivative.
$\alpha$ with the dimension of length-squared 
is the coupling constant of the Gauss-Bonnet term. 
We assume $\alpha \ge 0$ throughout this paper, as
motivated by string theory.
The gravitational equation derived from the action (\ref{action}) is
\begin{align}
{G^\mu}_{\nu} +\alpha {H}^\mu_{~~\nu} 
+\Lambda \delta^\mu_{~~\nu}= 
\kappa_n^2 {T}^\mu_{~~\nu}, \label{beq}
\end{align}
where 
\begin{align}
{G}_{\mu\nu}&:= R_{\mu\nu}-{1\over 2}g_{\mu\nu}R,\\
{H}_{\mu\nu}&:= 2\Bigl[RR_{\mu\nu}-2R_{\mu\alpha}
R^\alpha_{~\nu}-2R^{\alpha\beta}R_{\mu\alpha\nu\beta} \nonumber \\
&~~~~~~+R_{\mu}^{~\alpha\beta\gamma}R_{\nu\alpha\beta\gamma}\Bigr]
-{1\over 2}g_{\mu\nu}{L}_{\rm GB}
\end{align}
and 
$T_{\mu\nu}$
is the energy-momentum tensor of matter fields.
The field equations (\ref{beq}) contain up to the second derivatives 
of the metric and linear in that term.

Suppose the $n$-dimensional spacetime 
$({\ma M}^n, g_{\mu \nu })$ to be a warped product of an 
$(n-2)$-dimensional constant curvature space $(K^{n-2}, \gamma _{ij})$
and a two-dimensional orbit spacetime $(M^2, g_{ab})$ under 
the isometry of $(K^{n-2}, \gamma _{ij})$. Namely, the line element
is
\begin{align}
g_{\mu \nu }\D x^\mu \D x^\nu =g_{ab}(y)\D y^a\D y^b +r^2(y) \gamma _{ij}(z)
\D z^i\D z^j ,
\label{eq:ansatz}
\end{align} 
where
$a,b = 0, 1;~i,j = 2, ..., n-1$. 
Here $r$ is a scalar on $(M^2, g_{ab})$ with $r=0$ 
defining its boundary, and $\gamma_{ij}$ is the unit
metric on $(K^{n-2}, \gamma _{ij})$ with its sectional curvature $k = \pm 1, 0$. 
We assume that $({\ma M}^n, g_{\mu \nu})$ is strongly causal and 
$(K^{n-2}, \gamma _{ij})$ is compact. 
Since the rank-two symmetric tensors on the maximally symmetric space
are proportional to the metric tensor,
the symmetry of the background spacetime determines the structure of
the energy momentum tensor as
\begin{align}
T_{\mu \nu }\D x^\mu \D x^\nu =
T_{ab}(y)\D y^a \D y^b+p(y)r^2(y)\gamma 
_{ij}\D z^i\D z^j,
\end{align}
where $p(y)$ is a scalar function on $(M^2, g_{ab})$.

The generalized Misner-Sharp mass ~\cite{maeda2006b} 
is a scalar function on $(M^2, g_{ab})$ 
with the dimension of mass such that
\begin{align}
\label{qlm}
m &:= \frac{(n-2)V_{n-2}^k}{2\kappa_n^2}
\biggl\{-{\tilde \Lambda}r^{n-1}
+r^{n-3}[k-(D r)^2] \nonumber \\
&~~~~~~+{\tilde \alpha}r^{n-5}[k-(Dr)^2]^2 \biggl\},
\end{align}  
where ${\tilde \alpha} := (n-3)(n-4)\alpha$, 
${\tilde \Lambda} := 2\Lambda /[(n-1)(n-2)]$,
$D_a $ is a metric compatible linear connection on $(M^2, g_{ab})$
and $(Dr)^2:=g^{ab}(D_ar)(D_br)$.
$V_{n-2}^k$ is the area of the unit $(n-2)$-dimensional space of
constant curvature. 
The quasi-local mass is defined by the quasi-local geometrical quantity on the boundary of 
a spatial surface and dependent only on the metric and first derivatives.
It can be also derived by the locally conserved energy flux,
from which the quasi-local mass is recognized as a total amount of energy enclosing the
spatial surface \cite{mn2007}. 
The equations in the following analysis can be transcribed 
in a comprehensible form by using the quasi-local mass.
Physical properties of the quasi-local mass 
were elucidated in \cite{mn2007}, and partial results thereof will be used 
in the succeeding arguments.

In our analysis, it is suitable to write the 
line element in the double-null coordinates as
\begin{align}
\D s^2 = -2e^{-f(u,v)}\D u\D v
+r^2(u,v) \gamma_{ij}\D z^i\D z^j.
\end{align}
Null vectors $(\partial /\partial u)$ and $(\partial /\partial v)$ 
are taken to be future-pointing. 
The expansions of two independent future-directed 
radial null geodesics are defined as
\begin{align}
\theta_{+}&:=(n-2)r^{-1}r_{,v},\\
\theta_{-}&:=(n-2)r^{-1}r_{,u},
\end{align}  
where a comma denotes the partial derivative.
Note that the values of $\theta _+$ and $\theta _-$ are not the
geometrical invariants since the null coordinates $u$ and $v$ have
a residual rescaling freedom such as $u\to U=U(u), v\to V=V(v)$.
An invariant combination is $e^f \theta_+\theta_-$, which 
characterizes the trapping horizon as will be mentioned in the next section.
The function $r$, on the other hand, has a geometrical meaning
as an areal radius: the area of symmetric subspace
is given by $V^k_{n-2}r^{n-2}$.
Then, the quasi-local mass $m$ is expressed in a double-null form 
as
\begin{widetext}
\begin{align}
\label{qlm2}
m = \frac{(n-2)V_{n-2}^k}{2\kappa_n^2}r^{n-3}
\biggl[-{\tilde \Lambda}r^2+\left(k+\frac{2}{(n-2)^2} r^2e^{f}
\theta_{+}\theta_{-}\right)
+{\tilde \alpha}r^{-2}\left(k+\frac{2}{(n-2)^2} 
r^2e^{f}\theta_{+}\theta_{-}\right)^2\biggl],
\end{align}  
and the stress-energy tensor $T_{\mu\nu}$ as
\begin{align}
T_{\mu\nu}\D x^\mu \D x^\nu =
T_{uu}(u,v)\D u^2+2T_{uv}(u,v)\D u\D v
+T_{vv}(u,v)\D v^2+p(u,v)r^2 \gamma_{ij}\D z^i\D z^j.
\end{align}  
The governing field equations~(\ref{beq}) are
\begin{align}
&(r_{,uu}+f_{,u}r_{,u})\left[1+\frac{2{\tilde\alpha}}{r^2}
(k+2e^{f}r_{,u}r_{,v})\right]
=-\frac{\kappa_n^2}{n-2} r T_{uu}, \label{equation:uu} \\
&(r_{,vv}+f_{,v}r_{,v})\left[1+\frac{2{\tilde\alpha}}{r^2}
(k+2e^{f}r_{,u}r_{,v})\right]
=-\frac{\kappa_n^2}{n-2} r T_{vv}, \label{equation:vv} \\
&rr_{,uv}+(n-3)r_{,u}r_{,v}+\frac{n-3}{2}k e^{-f}+\frac{{\tilde\alpha}}{2r^2}
[(n-5)k^2e^{-f}+4rr_{,uv}
(k+2e^fr_{,u}r_{,v})+4(n-5)r_{,u}r_{,v}
(k+e^fr_{,u}r_{,v})] \nonumber \\
&~~~~~~-\frac{n-1}{2}{\tilde\Lambda}r^2e^{-f}
=\frac{\kappa_n^2}{n-2} r^2T_{uv},
 \label{equation:uv}\\
&r^2 f_{,uv}+2(n-3)r_{,u}r_{,v}
+k(n-3)e^{-f}-(n-4)rr_{,uv} \nonumber \\
&~~~~~~+\frac{2{\tilde\alpha}e^{-f}}{r^2}
\biggl[e^f(k+2e^fr_{,u}r_{,v})
\{r^2f_{,uv}-(n-8)rr_{,uv}\}
+2r^2e^{2f}\{(f_{,u}r_{,u}+r_{,uu})(f_{,v}r_{,v}+r_{,vv})-(r_{,uv})^2\} \nonumber \\
&~~~~~~+(n-5)(k+2e^fr_{,u}r_{,v})^2\biggl]=\kappa_n^2 r^2(T_{uv}+e^{-f}p). \label{equation:ij}
\end{align}  
\end{widetext}
The variation of $m$ is determined by these equations as
\begin{align}
m_{,v}&=
\frac{1}{n-2}V_{n-2}^ke^fr^{n-1}(T_{uv}\theta_+-T_{vv}\theta_-), \label{m_v} \\
m_{,u}&=
\frac{1}{n-2}V_{n-2}^ke^fr^{n-1}(T_{uv}\theta_- -T_{uu}\theta_+). \label{m_u} 
\end{align}  
These variation formulae are the same as those in general
relativity and therefore have several practical advantages.

In this paper, we do not specify the particular stress-energy tensor of matter fields.  
Alternatively, we impose energy conditions.
The null energy condition for the matter field implies
\begin{align}
T_{uu}\ge 0,~~~T_{vv} \ge 0, \label{nec}
\end{align}
while the dominant energy condition implies
\begin{align}
T_{uu} \ge 0,~~T_{vv}\ge 0,~~T_{uv}\ge 0, \label{dec}
\end{align}  
which assures that a causal observer measures the 
non-negative energy density and the energy flux is 
a future-directed causal vector.
The dominant energy condition implies the null energy condition,
but the converse is not true.

Unlike the general relativistic case, the quasi-local
mass~(\ref{qlm2}) is quadratic in $e^f\theta _+\theta _-$.
So they do not have one-to-one correspondence.
Solving Eq.~(\ref{qlm2}) inversely, we obtain
\begin{widetext}
\begin{align}
\label{trapping}
&\frac{2}{(n-2)^2} r^2 e^f\theta_{+}\theta_{-} = 
-k-\frac{r^2}{2{\tilde\alpha}}
\left(1\mp\sqrt{1+\frac{8\kappa_n^2{\tilde\alpha} m}
{(n-2)V^k_{n-2}r^{n-1}}+4{\tilde\alpha}{\tilde\Lambda}}\right).
\end{align}
\end{widetext}
There are two families of solutions corresponding to 
the sign in front of the square root in Eq.~(\ref{trapping}),
stemming from the quadratic curvature terms in the action.
We call the family having the minus (plus) sign 
the GR-branch (non-GR-branch) solution.
Note that the GR-branch solution has a general
relativistic limit as $\alpha \to 0$, 
\begin{align}
\label{trapping-gr}
\frac{2}{(n-2)^2} r^2 e^f\theta_{+}\theta_{-} = 
-k+\frac{2\kappa_n^2m}{(n-2)V^k_{n-2}r^{n-3}}+{\tilde\Lambda}r^2,
\end{align}  
but the non-GR branch does not.
Throughout this paper, the upper sign is used for the GR branch.

We also assume, in addition to 
$\alpha \ge 0$, the range of $\alpha $ as 
\begin{align}
\label{alphalambda}
1+4{\tilde\alpha}{\tilde\Lambda} \ge 0
\end{align}  
in order to avoid the zero-mass solution becoming unphysical. 
Eq. (\ref{alphalambda}) gives a restriction when $\Lambda $ is negative.
If the condition (\ref{alphalambda}) is satisfied, 
the (anti-)de Sitter space with effective cosmological constant
$\tilde \Lambda_{\rm eff}=(-1\pm\sqrt{1+4\tilde\alpha\tilde\Lambda})/(2\tilde\alpha)$
solves the vacuum field equations.

Under above conditions, it follows from Eq. (\ref{trapping}) that
the quasi-local mass has a non-positive lower bound
\begin{align}
\label{bc}
m \ge -\frac{(n-2)(1+4{\tilde\alpha}{\tilde\Lambda})V^k_{n-2}r^{n-1}}
{8\kappa_n^2{\tilde\alpha}}=:m_{\rm b}.
\end{align}
When the equality holds in Eq. (\ref{bc}), the two branches coincide.
We call the points where $m=m_{\rm b}$ holds {\it branch points}.

An immediate consequence of the variation formulae
is the constancy of the quasi-local mass in the absence of matter fields. 
If $1+4\tilde \alpha\tilde \Lambda \ne 0$ and $(Dr)^2\ne 0$, 
the general solution \cite{mn2007,Birkhoff}
is given by the generalized Boulware-Deser-Wheeler solution \cite{EGBBH}
\begin{align}
\D s^2 =-F(r)\D t^2+F^{-1}(r)\D r^2 +r^2 \gamma _{ij}
\D z^i\D z^j,
\label{BDW1}
\end{align}
where
\begin{align}
F(r):=k+\frac{r^2}{2\tilde \alpha }\left[1\mp
\sqrt{1+\frac{8\kappa _n^2\tilde \alpha m}{(n-2)V_{n-2}^kr^{n-1}}
+4\tilde \alpha \tilde \Lambda }\right].
\label{BDW2}
\end{align}
Analysis in~\cite{tm2005} 
provides a complete classification of the global 
structure of the generalized Boulware-Deser-Wheeler solution. 
(See~\cite{tm2005b} for the charged case.)
The event horizon in this vacuum spacetime is the simplest example 
of the trapping horizon discussed below.

\section{Trapping horizon and dynamical black hole}

The event horizon $H^+$ of a black hole is determined by the global structure of
a spacetime as $H^+=\dot J^-(\mas I^+)$, where $J^-$ and 
$\mas I^+$ denote a causal past and a future null infinity, respectively. 
Namely, one has to know the entire future of a spacetime to identify black-hole regions. 
However, it is rare to solve exactly the field equations 
due to its non-linearity, and therefore event horizons are of little 
use in identifying a black hole from a practical viewpoint. 
To overcome this difficulty, 
one may use a quasi-local notion of horizons, 
which is more easily handled
than the event horizon\footnote{
Even if the stationarity and the dominant energy condition are assumed,
the whole picture of the event horizon remains unclear 
in Einstein-Gauss-Bonnet gravity.
In general relativity, a powerful and useful theorem, called 
the rigidity theorem, is established 
\cite{arealaw,Hawking:1973uf,Chrusciel:1996bj,Hollands:2006rj}:
the event horizon in a stationary spacetime is a Killing horizon. 
The Killing horizon is totally geodesic,
and the surface gravity is constant over the horizon. 
Since the proof of rigidity heavily made use of Einstein equations, 
it has not been certain that the proof would proceed in parallel 
for the Einstein-Gauss-Bonnet gravity. 
}. 
The notion of trapping horizons was originally introduced 
by Hayward~\cite{hayward1994,hayward1996}. To begin with, 
we recapitulate the definitions.
We defer to \cite{Ashtekar:2003hk,Ashtekar:2004cn}
for the comparison with dynamical horizons.
\begin{dn}
\label{def:t-surface}
A {\it trapped (untrapped) surface} is a compact spatial 
$(n-2)$-surface with $\theta_{+}\theta_{-}>(<)0$.
\end{dn}
\begin{dn}
\label{def:t-region}
A {\it trapped (untrapped) region} is the union of all trapped (untrapped) surfaces.
\end{dn}
\begin{dn}
\label{def:m-sphere}
A {\it marginal surface} is an $(n-2)$-surface with $\theta_{+}\theta_{-}=0$.
\end{dn}
Without loss of generality, we set $\theta_{+}$ to be zero on a marginal
surface.
We also fix the orientation of the untrapped region such that
$\theta _+>0$ and $\theta_-<0$ in the hereafter. This means that 
$(\partial /\partial v)$ and $(\partial /\partial u)$ are 
pointing outward and inward, respectively.

\begin{dn}
\label{def:4-msphere}
A marginal surface is {\it future} if $\theta_-<0$, {\it past} if
$\theta_{-}>0$, {\it bifurcating} if $\theta_-=0$, {\it outer} if
$\theta_{+,u}<0$, {\it inner} if $\theta_{+,u}>0$ and {\it
degenerate} if $\theta_{+,u}=0$.
\end{dn}
\begin{dn}
\label{def:t-horizon}
A {\it trapping horizon} is the closure of a hypersurface 
foliated by future or past, outer or inner marginal surfaces.
\end{dn}

By definition, the notion of trapping horizons does not 
make any reference to the infinite future, nor the asymptotic
structure.
In contrast to event horizons, trapping horizons are meaningful 
even in the spatially compact spacetime.

Among all classes, the {\it future outer} trapping horizon is the most
relevant in the context of black holes~\cite{hayward1994,hayward1996}.
In this case, the definition expresses the idea that the ingoing 
null rays should be converging, $\theta_-<0$, and the outgoing null rays 
should be instantaneously parallel on the horizon, $\theta_+=0$, 
diverging just outside the horizon and converging just inside, $\theta_{+,u}<0$.

Since trapping horizons and event horizons are conceptually different,
one may suspect that there is no immediate relationship between them.
However, as is well known, a trapped region certainly arises in
the process of a black-hole formation from the 
gravitational collapse of a massive body. 
Now by the same arguments of Proposition 9.2.1 of 
\cite{Hawking:1973uf}, under the null convergence condition together
with the weak cosmic censorship,
we conclude $\theta _+\ge 0$ on $H^+$,
i.e., trapped regions cannot be seen from the future null infinity.
It then follows in general relativity that the trapping horizon coincides with or 
resides inside the event horizon under the null energy condition. 
Thus, they are mutually associated in physically reasonable circumstances.

The nature of trapping horizons in general relativity 
has been well appreciated together with energy conditions, 
which directly imply the null convergence condition. 
But in other theories of gravity, 
the relation between the convergence and energy conditions 
is not immediate via the field equations. 
For this reason, it is not apparent {\it a priori} that trapping horizons in
Einstein-Gauss-Bonnet gravity have the same properties
as those in general relativity. To elucidate this is the main 
goal of this section.

\subsection{Mass of the trapping horizon}

Because the concept of a trapping horizon is quasi-local, 
a quasi-local mass is adopted to evaluate the mass of a black hole.
The dynamical nature of a black hole was studied in \cite{hayward1996} 
in the four-dimensional spherically symmetric case in general relativity
without $\Lambda$, in which case a trapping horizon is succinctly
described by the Misner-Sharp mass.
Before moving on to the details, we review some of the basic properties of our quasi-local mass (\ref{qlm}). 
See \cite{mn2007} for the proof.

\begin{Prop} 
\label{th:asymptotics}
({\it Asymptotic behavior}.)
In the asymptotically flat spacetime,
$m$ reduces to the higher-dimensional Arnowitt-Deser-Misner 
(ADM) mass \cite{ADM} 
at spatial infinity.
\end{Prop}

\begin{Prop} 
\label{th:monotonicity}
({\it Monotonicity}.)
If the dominant energy condition holds, 
$m$ is non-decreasing (non-increasing) in any outgoing  
(ingoing) spacelike or null direction on an untrapped surface.
\end{Prop}

\begin{Prop} 
\label{th:positivity}
({\it Positivity}.)
If the dominant energy condition holds on an untrapped 
spacelike hypersurface with a regular center, 
then $m\ge 0$ holds there,
where the regular center denotes a central point $r=0$ with 
$k-(Dr)^2=O(r^2)$ in that neighborhood.
\end{Prop}

These properties support the well-posedness of the
quasi-local mass. The asymptotic value correctly denotes
the total energy, while monotonicity means that the mass contained 
within a spatial surface is non-decreasing outwardly. 
Positivity is not immediately manifest because of the negative contribution of
gravitational potential.
It should be stressed that since a regular center 
is always trapped for $k=-1$,
we cannot conclude the positivity of $m$ in this case, 
while the case where $k=1$ guarantees Proposition \ref{th:positivity}.
In the case where $k=0$, the assumption in the Proposition 
constraints on the metric form around the regular center.

Let us now look at the relation between the areal radius 
and the quasi-local mass of a trapping horizon.
From Eq.~(\ref{qlm2}), the mass of the trapping horizon 
with a radius $r=r_{\rm h}$ is
given by 
$m_{\rm h}(r_{\rm h})$, where
\begin{eqnarray}
m_{\rm h}(x) := \frac{(n-2)V_{n-2}^k}{2\kappa_n^2}x^{n-3}
\biggl(k+\frac{{\tilde \alpha k^2}}{x^2}-\tilde \Lambda x^2\biggl).
\label{THmass}
\end{eqnarray}  
In the special case with $1+4\tilde \alpha\tilde \Lambda=0$, we have
\begin{eqnarray}
m_{\rm h}(x)= \frac{(n-2)V_{n-2}^k}{8{\tilde\alpha}\kappa_n^2}
x^{n-5}(2{\tilde\alpha}k+x^2)^2\ge 0.
\label{THmass-s}
\end{eqnarray}

The succeeding two propositions are shown by direct 
calculations from the definition (\ref{trapping}).
These statements are not shared in the general relativistic case.

\begin{Prop}
\label{th:absenceTH}
({\it Absence of trapping horizons.}) 
An $(n-2)$-surface is necessarily untrapped, and trapping horizons are 
absent in the non-GR-branch solution for $k=0$ and $1$.
In the GR-branch (non-GR-branch) solution for $k=-1$ 
with $r^2<(>)2{\tilde\alpha}$, an $(n-2)$-surface is always trapped 
(untrapped), and trapping horizons are absent.
\end{Prop}

\begin{Prop}
\label{th:trapping}
({\it Trapping.}) 
In the GR-branch solution for $k=1,0$ and for $k=-1$ 
with $r^2\ge 2\tilde \alpha $
(In the non-GR-branch solution for $k=-1$ with 
$r^2\le 2\tilde \alpha$), 
an $(n-2)$-surface is trapped if and only if $m>(<) m_{\rm h}(r)$, 
marginal if and only if 
$m=m_{\rm h}(r)$ and untrapped if and only if 
$m<(>) m_{\rm h}(r)$.
\end{Prop}

Here we have implicitly assumed that the branch points are regular, 
so that trapping horizons can appear at the minimal (maximal) 
areal radius $r_{\rm h}=\sqrt{2\tilde \alpha }$ in the GR-branch 
(non-GR-branch) solutions for $k=-1$.
However, the branch points become singular in most cases, as we shall see
in Proposition \ref{b-sing}.
In the analysis below, 
we do not further consider trapping horizons with $r_{\rm h}=\sqrt{2\tilde \alpha }$ 
for $k=-1$.
Inasmuch as some equations become trivial at these points, propositions 
in the next subsections cannot be established. 
The exclusion of this special case is rather technical
than physically unrealizable
since our approach fails under the above special situation.

Now we turn to the task of inspecting the relation 
between $m_{\rm h}$ and $r_{\rm h}$, which can be 
completely understood from the result in~\cite{tm2005} for 
the generalized Boulware-Deser-Wheeler solution.
(The variable ${\tilde M}$ in~\cite{tm2005} is related to 
$m_{\rm h}$ as ${\tilde M} \equiv 2\kappa_n^2m_{\rm h}/[(n-2)V_{n-2}^k]$.)
The $m_{\rm h}$--$r_{\rm h}$ diagram is of great advantage in 
identifying the number of horizons and their types.

\begin{Prop}
\label{th:massgr}
({\it Horizon mass in general relativity.}) 
In general relativity, the mass of the trapping horizon 
$m_{\rm h}$ satisfies the inequalities in Table~\ref{table:massgr}, 
where $r_{\rm ex(GR)}:= [k(n-3)/\{(n-1){\tilde\Lambda}\}]^{1/2}$ 
and 
$m_{\rm ex(GR)}:=k(n-2)V^k_{n-2}r_{\rm ex}^{n-3}/[(n-1)\kappa _n^2]$.
\begin{table}[h]
\begin{center}
\caption{\label{table:massgr} Mass of the trapping horizon in general relativity.}
\begin{tabular}{l@{\qquad}c@{\qquad}c@{\qquad}c}
\hline \hline
  & $k=1$ & $k=0$ & $k=-1$   \\\hline
$\Lambda=0$ & $m_{\rm h}>0$ & $m_{\rm h}\equiv 0$ & $m_{\rm h}< 0$\\ \hline
$\Lambda>0$ & $m_{\rm h}\le m_{\rm ex(GR)}$ & $m_{\rm h}<0$ & $m_{\rm h}< 0$ \\ \hline
$\Lambda<0$ & $m_{\rm h}>0$ & $m_{\rm h}>0$ & $m_{\rm h}\ge m_{\rm ex(GR)}$ \\ 
\hline \hline
\end{tabular}
\end{center}
\end{table} 
\end{Prop}
\noindent
{\it Proof}.
See section III in~\cite{tm2005}.
\qed

\bigskip

The equality $m_{\rm h}=m_{\rm ex(GR)}$ attains when the two trapping horizons
are coincident, which produces the degenerate trapping horizon.

As discussed in~\cite{tm2005}, both the $n=5$ case as well as the
$1+4\tilde \alpha \tilde \Lambda =0$ case require
special treatment in Einstein-Gauss-Bonnet gravity.
This may be attributed to the fact that
$n=5$ is the lowest dimension in which the Gauss-Bonnet term
becomes nontrivial, and $1+4\tilde \alpha \tilde \Lambda =0$ is
the special combination of Lovelock coefficients, which yields
the Chern-Simons gravity for $n=5$ \cite{Banados:1993ur}.

\begin{widetext}
\begin{Prop}
\label{th:massgb}
({\it Horizon mass in Einstein-Gauss-Bonnet gravity.}) 
The mass of the trapping horizon $m_{\rm h}$ satisfies the inequalities 
in Table~\ref{table:mass6}, \ref{table:mass5} and \ref{table:mass-s} 
for $n \ge 6$ with $1+4{\tilde\alpha}{\tilde\Lambda}>0$, 
$n=5$ with $1+4{\tilde\alpha}{\tilde\Lambda}>0$ and 
$n\ge 5$ with $1+4{\tilde\alpha}{\tilde\Lambda}=0$, respectively, where
\begin{align}
m_{\rm ex}(x) &:= \frac{k(n-2)V^k_{n-2}}
{(n-1)\kappa _n^2}(x^2+2{\tilde\alpha}k)x^{n-5},\\
m_{\rm crit} &:= \frac{3\alpha V^1_{3}}{\kappa_5^2},\\
m_{\rm B} &:= -\frac{(2{\tilde\alpha})^{(n-3)/2}
(n-2)V^{-1}_{n-2}(1+4{\tilde\alpha}{\tilde\Lambda})}{4\kappa _n^2},\\
r_{\rm ex} &:= \biggl(\frac{-k(n-5){\tilde\alpha}}{n-3}\biggl)^{1/2},\\
r_{\rm ex(\pm)} &:= \biggl[-\frac{n-3}{2(n-1)
{\tilde\Lambda}}\biggl\{-k\pm |k|\sqrt{1+\frac{4{\tilde\alpha}
{\tilde\Lambda}(n-1)(n-5)}{(n-3)^2}}\biggl\}\biggl]^{1/2}.
\end{align}
\end{Prop}
\noindent
{\it Proof}.
See section IV in~\cite{tm2005}.
\qed
\begin{table}[h]
\begin{center}
\caption{\label{table:mass6} Mass of the trapping horizon in Einstein-Gauss-Bonnet 
gravity for $n \ge 6$ and $1+4{\tilde\alpha}{\tilde\Lambda}>0$.  Note that the inequality (\ref{bc}) may give a more severe constraint in the case with $k=-1$ and $\Lambda<0$ in the GR branch.}
\begin{tabular}{l@{\quad}|@{\quad}c@{\qquad}c@{\qquad}c@{\quad}|@{\quad}c@{\qquad}c@{\qquad}c}
\hline \hline
&\multicolumn{3}{c|@{\quad}}{GR branch} &
\multicolumn{3}{c}{non-GR branch} \\ 
& $k=1$ & $k=0$ & $k=-1$ & $k=1$ & $k=0$ & $k=-1$  \\\hline
$\Lambda=0$ & $m_{\rm h}>0$ & $m_{\rm h} \equiv 0$ & $m_{\rm h} < m_{\rm B}$ & n/a & n/a & $m_{\rm B} < m_{\rm h}\le m_{\rm ex}(r_{\rm ex})$\\ \hline
$\Lambda>0$ & $m_{\rm h}\le m_{\rm ex}(r_{\rm ex(-)})$ & $m_{\rm h}<0$ & $m_{\rm h}< m_{\rm B}$ & n/a & n/a & $m_{\rm B} < m_{\rm h}\le m_{\rm ex}(r_{\rm ex(-)})$\\ \hline
$\Lambda<0$ & $m_{\rm h}>0$ & $m_{\rm h}>0$ & $m_{\rm h}\ge m_{\rm ex}(r_{\rm ex(+)})$ & n/a & n/a & $m_{\rm
 B} < m_{\rm h}\le m_{\rm ex}(r_{\rm ex(-)})$\\ \hline \hline
\end{tabular}
\end{center}
\end{table} 
\begin{table}[h]
\begin{center}
\caption{\label{table:mass5} Mass of the trapping horizon in Einstein-Gauss-Bonnet 
gravity for $n=5$ and $1+4{\tilde\alpha}{\tilde\Lambda}>0$. Note that the inequality (\ref{bc}) may give a more severe constraint in the case with $k=-1$ and $\Lambda<0$ in the GR branch.}
\begin{tabular}{l@{\quad}|@{\quad}c@{\qquad}c@{\qquad}c@{\quad}|@{\quad}c@{\qquad}c@{\qquad}c}
\hline \hline
&\multicolumn{3}{c|@{\quad}}{GR branch} &
\multicolumn{3}{c}{non-GR branch} \\
  & $k=1$ & $k=0$ & $k=-1$ & $k=1$ & $k=0$ & $k=-1$  \\\hline
$\Lambda=0$ & $m_{\rm h}>m_{\rm crit}$ & $m_{\rm h} \equiv 0$ & $m_{\rm h}< m_{\rm B}$ & n/a & n/a & $m_{\rm B} < m_{\rm h}< m_{\rm crit}$\\ \hline
$\Lambda>0$ & $m_{\rm h}\le m_{\rm ex}(r_{\rm ex(-)})$ & $m_{\rm h}<0$ & $m_{\rm h}< m_{\rm B}$ & n/a & n/a & $m_{\rm B} < m_{\rm h}< m_{\rm crit}$\\ \hline
$\Lambda<0$ & $m_{\rm h}>m_{\rm crit}$ & $m_{\rm h}>0$ & $m_{\rm h}\ge m_{\rm ex}(r_{\rm ex(+)})$ & n/a & n/a & $m_{\rm B} < m_{\rm h}< m_{\rm crit}$\\ \hline\hline
\end{tabular}
\end{center}
\end{table} 
\begin{table}[h]
\begin{center}
\caption{\label{table:mass-s} Mass of the trapping horizon in Einstein-Gauss-Bonnet 
gravity for $n \ge 5$ and $1+4{\tilde\alpha}{\tilde\Lambda}=0$.}
\begin{tabular}{l@{\quad}|@{\quad}c@{\qquad}c@{\qquad}c@{\quad}|@{\quad}c@{\qquad}c@{\qquad}c}
\hline \hline
&\multicolumn{3}{c|@{\quad}}{GR branch} &
\multicolumn{3}{c}{non-GR branch} \\
  & $k=1$ & $k=0$ & $k=-1$ & $k=1$ & $k=0$ & $k=-1$  \\\hline
$n=5$ & $m_{\rm h} > m_{\rm crit}$ & $m_{\rm h}>0$ & $m_{\rm h} \ge 0$ & n/a & n/a & $0 \le  m_{\rm h}< m_{\rm crit}$\\ \hline
$n \ge 6$ & $m_{\rm h}>0$ & $m_{\rm h}>0$ & $m_{\rm h} \ge 0$ & n/a & n/a & $0 \le m_{\rm h} \le  m_{\rm ex}([2(n-5){\tilde\alpha}/(n-1)]^{1/2})$\\ \hline\hline
\end{tabular}
\end{center}
\end{table} 
\end{widetext}

\bigskip

The above propositions imply an upper or lower 
bound for the mass of the trapping horizon in some cases.
Although the class of trapping horizons is not specified here, 
it surely gives a constraint for the mass of a black hole defined by a trapping horizon.

Next, we show the following mass inequality in Einstein-Gauss-Bonnet gravity.
\begin{Prop}
\label{th:mass}
({\it Mass inequality.}) 
If the dominant energy condition holds, then $m \ge (\le) m_{\rm h}(r_{\rm h})$ 
holds in the GR branch (non-GR branch) on an untrapped spacelike
hypersurface of which the inner boundary is a marginally trapped surface with radius $r_{\rm h}$.
\end{Prop}
\noindent
{\it Proof}.
By Proposition \ref{th:monotonicity}, we have 
$m \ge m|_{r=r_{\rm h}} \equiv m_{\rm h}(r_{\rm h})$ 
on the untrapped spacelike hypersurface.
\qed

\bigskip

The positivity of $m$ in the untrapped region with a regular center 
was shown in Proposition~\ref{th:positivity}.
On the other hand, Proposition~\ref{th:mass} claims that in the GR branch
there may be a more severe lower bound on $m$ on the untrapped hypersurface 
of which the inner boundary is a marginally trapped surface.
For $k=1$ and $\Lambda \le 0$, for example, there is a positive lower
bound on $m$.
If there is a black or white hole with area $\ma A_{n-2}r_{\rm h}^{n-2}$,
where $\ma A_{n-2} (:=V^1_{n-2})$ is the area of a unit $(n-2)$-sphere, 
the mass-energy measured outside the hole satisfies an isoperimetric inequality
\begin{align}
m\ge \frac{(n-2)\ma A_{n-2}}{2\kappa _n^2}r_{\rm h}^{n-3}
\biggl(1+\frac{\tilde{\alpha }}{r_{\rm h}^2}-\tilde{\Lambda }
r_{\rm h}^2 \biggl)
=: m_{\rm irr}.
\label{penrosein}
\end{align}
$m_{\rm irr}(>0)$ represents the minimal mass of a black hole or white hole,
corresponding to the irreducible mass.
For $n=4, k=1, \Lambda =0$, the above inequality
becomes $\sqrt{(4\pi r_{\rm h}^2)/16\pi}\le G_4m$, 
which is comparable to the Penrose inequality
\cite{Penrose:1969pc,Gibbons1972,Bray:2003ns}.
If the untrapped surface extends to spacelike infinity
in the asymptotically flat case, 
Proposition \ref{th:asymptotics} and \ref{th:mass} 
prove the special case of the positive mass theorem for black holes
\cite{Gibbons:1982jg}.
On the other hand, when $\Lambda$ is positive, $m_{\rm h}(r_{\rm h})$ 
may be negative for some $r_{\rm h}>0$, and then 
this result does not give a stronger lower bound on $m$.

\subsection{Properties of the trapping horizon}
In this subsection, we investigate properties of the trapping horizon.
Among all classes, a future outer trapping horizon defines 
a dynamical black hole and is particularly important.

In four-dimensional general relativity, the topology 
of an outer trapping horizon is restricted to either a two-sphere or a two-torus
if we assume $\Lambda \ge 0$ and the dominant energy condition
\cite{hayward1994}, which is the correspondent of 
Hawking's topology theorem\footnote{
Event horizons with toroidal topology cannot be realized in four dimensions
if the null convergence condition, 
$\mathscr I \simeq \mathbb R \times S^2$ (this condition automatically
holds if the spacetime is asymptotically flat or anti-de Sitter) and the weak cosmic censorship
are assumed,
as a consequence of the topological censorship
\cite{TC}.}\cite{arealaw,Hawking:1973uf}. 
A major difficulty encountered in Einstein-Gauss-Bonnet gravity is whether
we can draw appropriate information on spacetime curvatures just from the energy condition.
Under the present spacetime ansatz (\ref{eq:ansatz}), however, it is rather straightforward.
\begin{Prop}
\label{th:topology}
({\it Topology.}) 
An outer trapping horizon must have a
topology of non-negative curvature in the GR branch 
if $\Lambda \ge 0$ and the dominant 
energy condition is satisfied.
\end{Prop}
\noindent
{\it Proof}.
Evaluation of Eq. (\ref{equation:uv}) on a trapping horizon gives
\begin{align}
&
\frac{(n-2)k}{2r_{\rm h}^2}[n-3+(n-5)\tilde \alpha kr^{-2}_{\rm h}]
\nonumber \\
&~~~~~
=\kappa _n^2e^fT_{uv}+\Lambda -e^f\theta _{+,u}
\left(1+\frac{2\tilde \alpha k}{r^2_{\rm h}}\right).
\label{Tuveq1}
\end{align}
Now let
$\Lambda \ge 0$ and the dominant energy condition
be assumed; it then follows from Proposition~\ref{th:absenceTH}
that the right-hand-side of Eq.~(\ref{Tuveq1}) is nonnegative 
for an outer trapping horizon in the GR branch. 
For $k=-1$, since we have $r^2_{\rm h}> 2\tilde \alpha $ by Proposition
\ref{th:absenceTH}, the left-hand-side of Eq.~(\ref{Tuveq1}) becomes
negative, which yields inconsistency.
Thus, only the case where $k=1$ or $0$ is possible. 
\qed

\bigskip

Note that an outer trapping horizon with the $k=0$ topology 
can appear if and only if $\Lambda=0$ 
and $e^f =0$ on the trapping horizon.
These black holes are considered to be non-generic and 
therefore rarely to develop because they occur under highly restrictive conditions.

Unfortunately, 
there is no sign control of $k$ when $\Lambda <0$ in the GR branch, in which case
various topology is allowed. 
In the non-GR branch, any class of trapping horizons must have a
topology of negative curvature, as shown in
Proposition~\ref{th:absenceTH}, 
irrespective of the energy conditions and the sign of $\Lambda$.

The following lemma is used in the proof for later propositions. 
Observe that trapping horizons coinciding with 
the branch points are excluded from our consideration. 
\begin{lm}
\label{lm:+v}
If the null energy condition holds, $\theta_{+,v}\le (\ge) 0$ 
is satisfied on the trapping horizon in the GR (non-GR) branch .
\end{lm}
{\it Proof}. 
From Eq.~(\ref{equation:vv}), we obtain
\begin{eqnarray} 
2\theta_{+,v}\biggl(1+\frac{2k{\tilde\alpha}}{r^2_{\rm h}}\biggl)
=-\kappa^2_nT_{vv}
\label{Tvv}
\end{eqnarray}
on the trapping horizon, which gives
$\theta_{+,v}(r^2_{\rm h}+2k{\tilde\alpha})\le 0$
 by the null energy condition. 
Consequently, the above lemma follows from Proposition~\ref{th:absenceTH}.
\qed

\bigskip
Now let 
$
\xi^\mu(\partial/\partial x^\mu)=
\xi^u(\partial/\partial u)+\xi^v(\partial/\partial v)
$
be the generator of the trapping horizon. 
Since the trapping horizon is foliated by the marginal surfaces, 
\begin{align}
\mathscr L_\xi \theta_{+}=\theta_{+,v}\xi^v+\theta_{+,u}\xi^u=0
\label{expansionderiv}
\end{align}
holds on the trapping horizon.
If the trapping horizon is null ($\xi ^u=0$),
$\theta _{+,v}=0$ is concluded. Then Eq. (\ref{Tvv}) signifies that
there is no energy inflow $T_{vv}=0$ across the null 
trapping horizon, irrespective of energy conditions.

\begin{Prop}
({\it Signature law.})
\label{prop:sig} 
Under the null energy condition, an outer (inner) 
trapping horizon in the GR branch is non-timelike (non-spacelike), 
while it is non-spacelike (non-timelike) in the non-GR branch.\footnote{
By ``trapping horizon is non-timelike'' we mean that 
the generator of the trapping horizon $\xi ^\mu $, everywhere orthogonal to a 
foliation of marginal surfaces and preserving the foliation, is non-timelike.}
\end{Prop}
\vspace{5mm}
{\it Proof}. 
From Eq. (\ref{expansionderiv}), we have  
\begin{eqnarray} 
\xi ^u=-\frac{\theta_{+,v}}{\theta_{+,u}}\xi ^v,
\end{eqnarray}
on the trapping horizon.
Thus, by Lemma~\ref{lm:+v}, 
$\xi ^v\xi ^u \le (\ge)0 $
is satisfied on the outer (inner) trapping horizon in the GR branch, 
while $\xi ^v\xi ^u \ge (\le)0 $
is satisfied on the outer (inner) trapping horizon in the non-GR branch. 
 \qed

\bigskip

\begin{Prop}
({\it Trapped side.})
\label{prop:futureinner}
Let the null energy condition be assumed.
Then, the outside (inside) region of a future inner trapping 
horizon is trapped (untrapped) in the GR
 branch, and the outside (inside) region of a future outer trapping 
horizon is untrapped (trapped) in the non-GR
 branch. 
To the contrary, 
the future (past) domain of a future outer trapping horizon 
is trapped (untrapped) in the GR branch, and the 
future (past) domain of future inner trapping horizon is untrapped
 (trapped) in the non-GR branch.
\end{Prop}
\noindent
{\it Proof}. 
Along a vector 
$s^\mu(\partial/\partial x^\mu)=s^v(\partial/\partial
v)+s^u(\partial/\partial u)$, 
we obtain 
$\mathscr L_s (\theta_{+}\theta_{-})=
\theta_{-}(s^v\theta_{+,v}+s^u\theta_{+,u})$ 
on a trapping horizon.
First, let us take $s^\mu $ to be an outgoing spatial vector, 
where $s^v>0$ and $s^u<0$ are satisfied.
Then, by Lemma~\ref{lm:+v}, $\mathscr L_s (\theta_{+}\theta_{-})>0$
holds on a future inner trapping horizon in the GR branch, 
while $\mathscr L_s (\theta_{+}\theta_{-})<0$ holds on 
a future outer trapping horizon in the non-GR branch.
Next let $s^\mu$ be a future-directed timelike
vector, 
where $s^v>0$ and $s^u>0$ are satisfied.
Then, by Lemma~\ref{lm:+v}, $\mathscr L_s (\theta_{+}\theta_{-})>0$ 
holds on a future outer trapping horizon in the GR branch, 
while $\mathscr L_s (\theta_{+}\theta_{-})<0$ holds on 
a future inner trapping horizon in the non-GR branch.
\qed

\bigskip

Propositions~\ref{prop:sig} and \ref{prop:futureinner} mean that 
a future outer trapping horizon in the GR branch is a one-way membrane 
being matched to the concept of a black hole as a region of no escape.
On the other hand, the non-GR branch is diametrically opposed.
One might hope that a future outer trapping horizon in the non-GR branch 
also deserves to be called a black-hole horizon since Proposition~\ref{prop:futureinner} 
shows that it is an inner boundary of untrapped surfaces.
However, Proposition~\ref{prop:sig} claims that it does not capture 
the idea that a black hole is a one-way membrane. 
A light ray emanating from a point on a future outer trapping horizon 
can propagate into both sides of it since it can be timelike. 
Then there naturally arises a question: what causes such 
an antithetical and pathological behavior in the non-GR branch?

We digress here to discuss this issue further.
The two branches stem from the quadratic terms 
in curvature and are confluent at the branch points. 
The plus-minus sign in Eq. (\ref{trapping}) makes the respective 
branches quite different.
The following lemma answers the above question.
\begin{lm}
\label{ncc}
If $T_{\mu \nu }k^\mu k^\nu \ge 0$ is satisfied for a radial null vector $k^\mu $, 
$R_{\mu \nu }k^\mu k^\nu \ge (\le)0$
holds in the GR (non-GR) branch.
\end{lm}
{\it Proof.}
Using Eqs.~(\ref{equation:uu})--(\ref{equation:ij}) 
together with the expressions of the Ricci tensors, we obtain
\begin{align}
T_{\mu\nu}k^\mu k^\nu =&\frac{1}{\kappa_n^2}R_{\mu\nu}k^\mu k^\nu
\biggl[1+\frac{2{\tilde\alpha}}{r^2}(k+2e^fr_{,u}r_{,v})\biggl] \nonumber \\
&+\frac{8\tilde \alpha e^{-f}}{\kappa_n^2 r^4}
k^uk^v\biggl[\frac{r^4e^{2f}}{(n-2)^2}R_{uu}R_{vv} \nonumber \\
&-(k+2e^fr_{,u}r_{,v}-e^frr_{,uv})^2\biggl]
\label{focus-rel}
\end{align}
for a general null vector $k^\mu$.
For a radial null vector, where 
$k^\mu(\partial/\partial x^\mu)=k^u(\partial/\partial u)$
 or $k^v(\partial/\partial v)$, Eqs. (\ref{focus-rel}) and (\ref{trapping}) combine to give
\begin{align}
\pm R_{\mu\nu}k^\mu k^\nu\sqrt{1+\frac{8\kappa_n^2{\tilde\alpha} m}
{(n-2)V^k_{n-2}r^{n-1}}+4{\tilde\alpha}{\tilde\Lambda}}
=\kappa _n^2T_{\mu\nu}k^\mu k^\nu.
\label{Rvv}
\end{align}
\qed
\bigskip

This lemma shows that the null convergence condition 
$R_{\mu\nu}k^\mu k^\nu \ge 0$ fails in the non-GR branch 
if the null energy condition is {\it strictly} satisfied $T_{\mu\nu}k^\mu k^\nu > 0$.
The Vaidya-type radiating solution gives such an example
\cite{maeda2006,Kobayashi:2005ch}.
It signals that solutions in the non-GR branch behave badly 
under the null energy condition, since properties of the geometry are
determined not by energy conditions but by the convergence condition,
as seen in the Raychaudhuri equation. In the non-GR-branch solution, 
gravity effectively acts repulsively for
the positive energy particles. Lemma \ref{ncc} is most convincing
to account for the peculiarity of the non-GR-branch solution.

In this paper, trapping horizons coinciding with the branch points are excluded from our considerations.
This decision is strongly supported by the following proposition.
\begin{Prop}
\label{b-sing}
({\it Branch singularity.}) 
If the null energy condition is strictly satisfied at least for a
radial null vector, the branch points are curvature singularities.
\end{Prop}
\bigskip
{\it Proof}. 
Let $k^\mu=k^a(\partial /\partial x^a)^\mu$ be a radial null vector.
It then follows from Eq.~(\ref{Rvv}) that we have $R_{\mu\nu}k^\mu k^\nu\to \pm\infty$ 
at the branch points, where the inside of the square-root in Eq.~(\ref{Rvv}) vanishes.
\qed

\bigskip

In Proposition~\ref{b-sing}, the null energy condition must be
strictly satisfied for a radial null vector; however, 
the appearance of a singularity is not necessarily due to the presence
of matter fields.
Even in the vacuum case, the generalized Boulware-Deser-Wheeler solution with 
$1+4\tilde\alpha \tilde\Lambda >0$ has a 
branch singularity where the Kretschmann scalar 
$R_{\mu \nu \rho \sigma }R^{\mu\nu \rho \sigma }$ diverges. 
As it now stands, we have no definite answer for how generic the curvature
singularity is when $T_{ab}=0$ holds at the branch points. 
We leave this to future investigation.

Here we also show the following proposition as another consequence 
of Lemma~\ref{ncc}, claiming that, as in general relativity, 
caustics develop in a congruence of a radial null geodesic 
in the GR branch if the convergence occurs anywhere.
\begin{Prop}
\label{caustic}
({\it Caustics.}) 
Let $k^\mu$ be the tangent to an affinely parametrized 
radial null geodesic, and let the null energy condition hold.
If the expansion $\theta:=\nabla_\mu k^\mu$ takes the negative value $\theta_0$ 
at any point on a geodesic in the congruence, 
then $\theta \to -\infty$ along that geodesic within the affine length
$\lambda \le (n-2)/|\theta_0|$ in the GR branch, provided
that the geodesic is extended to this parameter value.
\end{Prop}
\bigskip
{\it Proof}. 
The Raychaudhuri equation for an affinely parametrized radial null
geodesic with tangent $k^\mu=k^a(\partial/\partial x^a)^\mu$ is written as
\begin{align}
\frac{\D \theta}{\D \lambda} =  -\frac{1}{n-2}\theta^2-R_{\mu\nu}k^\mu k^\nu,
\label{ray}
\end{align}
where $\lambda$ is an affine parameter.
By Lemma~\ref{ncc}, the above equation gives 
\begin{align}
\frac{\D \theta}{\D \lambda}+\frac{1}{n-2}\theta^2 \le 0
\end{align}
in the GR branch, implying 
\begin{align}
\frac{\D}{\D \lambda}(\theta)^{-1} \ge \frac{1}{n-2}
\end{align}
and hence
\begin{align}
\theta(\lambda) \le \frac{(n-2)\theta_0}{(n-2)+\lambda\theta_0}, \label{ray2}
\end{align}
where $\theta_0$ is the initial value of $\theta$.
If $\theta_0<0$, then Eq.~(\ref{ray2}) gives $\theta \to -\infty$ 
within an affine parameter $\lambda \le (n-2)/|\theta_0|$, provided that
the geodesic can be extended that far.
\qed

\bigskip

Let us now turn to a discussion of the area law
of a trapping horizon.
The area of a black-hole event horizon 
is non-decreasing into the future 
under the null convergence condition~\cite{arealaw},
which is transcribed into the null energy condition, 
via Einstein equations, in general relativity.
Then, how about the trapping horizon? It should be emphasized that
the proof of Hawking's area theorem relies on the Raychaudhuri equation along the
null geodesic generator of the event horizon.
We extrapolate that
the area theorem fails for the future outer trapping horizon
in the non-GR branch. The next proposition shows that 
this is indeed the case.

\begin{Prop}
\label{arealaw}
({\it Area law.}) 
Under the null energy condition, the area of a future outer 
(inner) trapping horizon is non-decreasing
 (non-increasing) along the generator of the trapping horizon 
in the GR branch, while it is non-increasing (non-decreasing) 
in the non-GR branch.
\end{Prop}
\bigskip
{\it Proof}. 
The derivation of the area 
\begin{align}
 A(r):= V_{n-2}^k r^{n-2},
\label{area}
\end{align}
along the generator of a trapping horizon $\xi^\mu$ is given by
\begin{align}
\mathscr L_ \xi A & =(n-2)r^{n-3}V_{n-2}^k(r_{,u}\xi^u+r_{,v}\xi^v),
\nonumber \\
&=r^{n-2}_{\rm h}V_{n-2}^k\theta_{-}\xi^u,
\end{align}  
where the second equality is evaluated on the the trapping horizon.
Here we fix the orientation such that $\xi^v>0$, which guarantees 
the non-spacelike (spacelike) trapping horizon to be future-directed (outgoing).
Then, we obtain $\mathscr L _\xi A \ge(\le) 0$ on the future outer 
(inner) trapping horizon in the GR branch and on the future inner 
(outer) trapping horizon in the non-GR branch by Proposition~\ref{prop:sig}. 
\qed

\bigskip

Here we note that the above area law takes the meaning of ``time
evolution'' only when the trapping horizon is non-spacelike. 
By Proposition~\ref{prop:sig}, such a trapping horizon 
is inner and outer in the GR and non-GR branches, respectively.
On the other hand, if the trapping horizon is null, 
an outer trapping horizon in the GR branch and an inner 
trapping horizon in the non-GR branch have this property.
When the generator of the trapping horizon is spacelike, the theorem simply
says that it is outward pointing.

\begin{figure}[htbp]
\begin{center}
\includegraphics[width=0.9\hsize]
{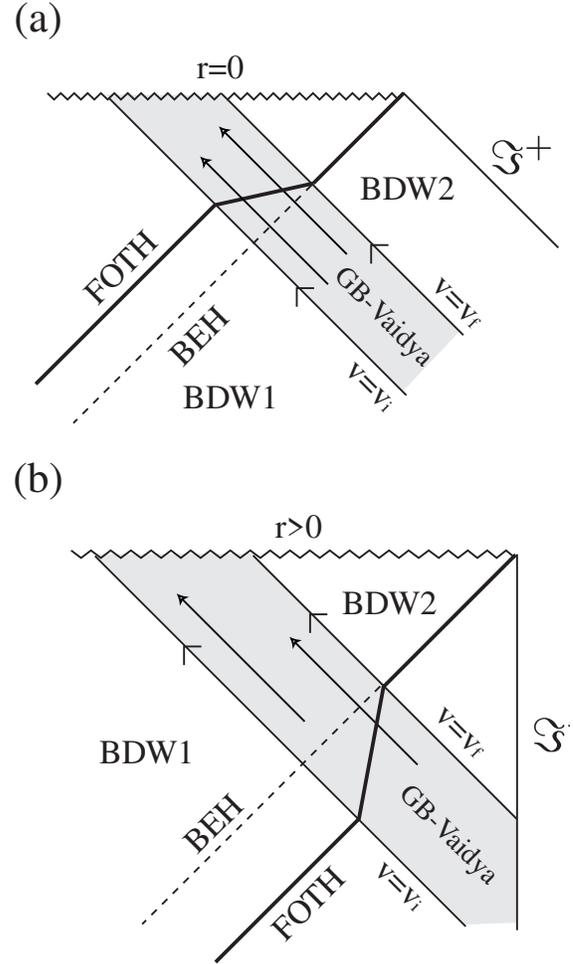}
\caption{\label{Penrose}
A portion of the Penrose diagram representing the transition from 
a generalized Boulware-Deser-Wheeler black hole~\cite{EGBBH} with mass  $m_1$ (BDW1) 
to another generalized Boulware-Deser-Wheeler black hole with mass $m_2(>m_1)$ (BDW2) 
by an incident null dust fluid with positive energy density in the 
(a) GR branch (where $k=1$, $\Lambda=0$ and $0<m_1<m_2$) and the
(b) non-GR branch (where $k=-1$ and $m_{\rm B}<m_1<m_2<0$).
BEH (a dashed line) and FOTH (a thick solid line) mean a black-hole
 event horizon and a future outer trapping horizon, respectively.
Here BDW1 spacetime for $v<v_{\rm i}$ is joined to BDW2 spacetime 
for $v>v_{\rm f}$ by way of the Vaidya-type (GB-Vaidya) spacetime~\cite{maeda2006,Kobayashi:2005ch}.
The zigzag line corresponds to a curvature singularity.
The future outer trapping horizon in the null-dust region is spacelike (timelike) in the GR (non-GR) branch.
}
\end{center}
\end{figure}

In closing this subsection, we give an example of dynamical spacetimes 
containing a future outer trapping horizon in Fig.~\ref{Penrose}.
It represents the transition from a generalized Boulware-Deser-Wheeler 
black hole with mass $m_1$ to another generalized Boulware-Deser-Wheeler 
black hole with mass $m_2(>m_1)$ by an incident null dust fluid with positive energy density.

\subsection{Black-hole dynamics}
\label{subsec:BHD}

Black-hole thermodynamics is now established for a stationary spacetime
in general relativity (see e.g, \cite{wald,wald2}).
Even in non-stationary and highly dynamical situations, 
analogous laws hold for a trapping horizon in the 
four-dimensional general relativistic
case \cite{hayward1994,Hayward:1994yy,Hayward:2004dv,Hayward:2004fz,hayward1996}.
It may be tempting to hope that similar results go through in other
theories of gravity.
The aim of this subsection is to address the issue of black-hole dynamics  
in Einstein-Gauss-Bonnet gravity.

A well-defined mass should satisfy the first law, representing
the energy conservation. This is validated also for the quasi-local mass
(\ref{qlm}) \cite{mn2007}.
We define a scalar 
\begin{align}
P:=-\frac 12 {T^a}_a,
\end{align}
and a vector
\begin{align}
\psi ^a :={T^a}_bD ^b r +PD ^a r
\end{align}
on $(M^2, g_{ab})$,
where the contraction is taken over on the two-dimensional orbit space.
It is also convenient to use the areal volume 
\begin{align}
V:=\frac{V_{n-2}^k}{n-1}r^{n-1}
\label{volume}
\end{align}
satisfying $D_a V=AD _ar$, where $A$ is defined by
Eq. (\ref{area}). 
By using the field equations (see 
appendix of \cite{mn2007}), 
we obtain
\begin{align}
\D  m =A\psi_a\D x^a  +P\D V.
\label{1stlaw1}
\end{align}
This is the unified first law \cite{hayward1998}
corresponding to the energy balance law, which reduces to 
the variation formulae (\ref{m_v}) and (\ref{m_u}) in the double null coordinates.
In the form (\ref{1stlaw1}), the physical meaning of each term is more 
readily recognizable.
The first term represents an energy flux, 
while the second an external work \cite{hayward1998,ashworth1999,Hayward:1998ee}. 
Assuming the dominant energy condition, 
we have $P\ge 0$. 
$\psi^a$ corresponds to the quasi-localization of 
the Bondi-Sachs energy loss \cite{Bondi}, but its interpretation in odd spacetime
dimensions remains unclear \cite{hollands2005,hollands2004}.
We will not discuss this issue because it is beyond the scope of
the present paper.
We anticipate that the evaluation 
of the unified first law (\ref{1stlaw1})
on the trapping horizon gives the first law of black-hole mechanics.
To this end, we must read off various ``thermodynamical quantities.''

We here follow in the footsteps of Killing horizons, 
which have been fully studied in the literature and are now
well established. 
Let $\zeta ^\mu $ be a horizon-generating Killing field of a Killing horizon.
The surface gravity, $\kappa $, of a Killing horizon is defined by
\cite{wald2, wald}
\begin{align}
\zeta ^\nu \nabla _\nu \zeta ^\mu  =\kappa \zeta ^\mu ,
\label{surfacegrakilling}
\end{align}
where the equality is evaluated on the Killing horizon.
$\kappa $ measures the non-affinity of the Killing field
and remains constant over the horizon if it has a regular bifurcation
surface \cite{Racz:1992bp}.
What plays the role of a horizon-generating Killing field 
for a trapping horizon? 
We shall embrace the generalized Kodama vector (simply the Kodama
vector, hereafter) 
as a substitute \cite{kodama1980,ms2004}, defined by
\begin{align}
K^a =-\epsilon ^{a b}D_b r, 
\label{kodamavec}
\end{align}
where $\epsilon _{ab}$ is a volume element of 
$(M^2, g_{ab})$. 
As shown in \cite{mn2007}, the Kodama vector is intimately related to
our definition of quasi-local mass. It follows immediately 
by definition that
\begin{align}
K^aK_a=-(D r)^2,
\label{ksquare}
\end{align}
so that the Kodama vector generates a preferred time evolution vector
field in the untrapped region. On the trapping horizon, 
$K^a$ becomes null and is given by $K^a=D^a r$.
A naive prescription for defining the surface gravity of a trapping horizon
is to replace 
$\kappa $ by $\kappa _{\rm TH}$ and $\zeta ^\mu$ by $K^a$
in Eq. (\ref{surfacegrakilling}) and to valuate the equality at the 
trapping horizon.
Simple calculations show
\begin{align}
K^bD _bK_a &=(D^2 r)D_a r-(D^b r)D_bD_a r, 
\label{derivk1}\\
K^bD_{(b}K_{a)} &=\frac12 (D^2 r)D _a r-(D^br)D_bD_a r,\nonumber \\
&=\frac{r\kappa _n^2}{n-2}\left(1+\frac{2\tilde \alpha }{r^2}
[k-(Dr)^2]\right)^{-1}\psi_a,
\label{derivk2}
\end{align}
where we have used two-dimensional identities
$\epsilon_{ab}\epsilon^{cd}\equiv -2{\delta^c}_{[a}{\delta^d}_{b]}$
and 
\begin{align}
&
\biggl(D_a D_b r-\frac 12 g_{ab}D^2 r \biggr)D^2 r \nonumber \\
& \qquad 
\equiv (D_aD^c r)(D_bD_cr) -\frac 12 g_{ab}(D_cD_d r)(D^cD^d r)
\label{S=0}
\end{align}
to derive these equations.
Note that at this stage equalities in these equations are
not restricted on the trapping horizon.
Eq. (\ref{derivk2}) reveals that $\psi^a$ vanishes if 
$K^a $ is a Killing vector on ($M^2, g_{ab}$), implying that
$K^\mu =K^a(\partial /\partial x^a)^\mu $ is a hypersurface-orthogonal
Killing vector on ($\ma M^n, g_{\mu \nu }$). This fact also lends 
support to the physical interpretation of $\psi^a$.
Since $\psi_aK^a=T_{ab}K^aK^b$ on the trapping horizon
where $D^a r=K^a$ holds,
$\psi^a$ is not in general proportional to $K^a$ in a dynamical setting.
Then the surface gravity of a trapping horizon should be defined by
$K^bD_{[b}K_{a]}=\kappa _{\rm TH}K_a$. Thus we have
\begin{align}
\kappa_{\rm TH} =\frac12 D^2 r=-\frac 12 \epsilon ^{ab}D_aK_b,
\label{surfacegravity}
\end{align}
where the evaluation is performed on the trapping horizon.
Note that Eq. (\ref{surfacegravity}) is expressed in a purely 
geometrical way and confirms that the surface gravity vanishes
for a degenerate trapping horizon. Note also that
even along the Kodama vector, the surface gravity is not constant
in general,
which reflects the non-equilibrium situation.

After some manipulations,
we can rewrite the unified first law (\ref{1stlaw1}) as
\begin{align}
A\psi _a =&\frac{D^2 r}{2\kappa _n^2}\left\{D_a A+2(n-2)
\tilde \alpha V_{n-2}^k
r^{n-5}[k-(Dr)^2]D_ar \right\} \nonumber \\
&+\frac{(n-2)\tilde \alpha V_{n-2}^kr^{n-6}}{\kappa _n^2 }
\left\{[k-(Dr)^2]^2-k^2\right\}D_a r\nonumber \\
& +r^{n-3}D_a\left[\frac{m}{r^{n-3}}+\frac{(n-2)V_{n-2}^k}{2\kappa _n^2}
\left(\tilde \Lambda r^2-\frac{\tilde \alpha k^2}{r^2}\right)\right].
\label{1stlaw2}
\end{align}
The second and third
terms on the right-hand-side of Eq. (\ref{1stlaw2}) vanish
along the trapping horizon (see Eq. (\ref{THmass})), where $(Dr)^2=0$.
Thus we obtain the desired first law for a trapping horizon
\begin{align}
A\xi ^a\psi_a 
=\frac{\kappa_{\rm TH} }{\kappa _n^2}\xi ^a D_a\left[
A\left( 1+\frac{2 (n-2)\tilde \alpha k}{(n-4)r^2}
\right)\right],
\label{1stTH}
\end{align}
where $\xi ^a$ denotes the generator of a trapping horizon.
Since the unified first law (\ref{1stlaw1}) gives
\begin{align}
\mathscr L_ \xi  m =A\psi_a \xi^a  +P\mathscr L_ \xi V,
\label{1stlawTH}
\end{align}
the first term on the right side is regarded as 
$A\psi_a\xi^a \equiv T_{\rm TH}\mathscr L_\xi S_{\rm TH}$, 
where $T_{\rm TH}$ and $S_{\rm TH}$ are the temperature and entropy 
of a trapping horizon, respectively.
Then, by identifying the temperature with 
$T_{\rm TH}:=  \kappa_{\rm TH} /(2\pi)$
as for a Killing horizon, the entropy of a 
trapping horizon $S_{\rm TH}$ is obtained from Eq.~(\ref{1stTH}) as
\begin{align}
S_{\rm TH} :=&\frac{2\pi A(r_{\rm h})}{\kappa _n^2}
\left[
1+\frac{2 (n-2)\tilde \alpha k}{(n-4)r_{\rm h}^2}\right], 
\nonumber \\
=&
\frac{V_{n-2}^k r_{\rm h}^{n-2}}{4 G_n}\left[
1+\frac{2 (n-2)(n-3)\alpha k}{r_{\rm h}^2}\right].
\label{entropy}
\end{align}
This coincides with Iyer and Wald's definition of 
dynamical black-hole entropy \cite{iyerwald1994,lb2007}, which
has several plausible properties among other things. 
Their entropy is independent of the potential ambiguity of the Lagrangian 
and associated with a Noether charge. Moreover, it agrees with
a non-stationary perturbation of the entropy of a stationary
black hole and reduces to the entropy of a stationary black hole
in the stationary case.  
For the generalized Boulware-Deser-Wheeler solution (\ref{BDW1}), the black-hole 
entropy is given by the replacement of $r_{\rm h}$ by $r_+$ in
Eq. (\ref{entropy}) \cite{jacobson1993,Cho:2002hq,clunan2004,Kofinas:2006hr},
where $r_+$ is the root of $F(r_+)=0$ in Eq. (\ref{BDW2}) 
denoting the location of the black-hole event horizon. (See also \cite{Cvetic:2001bk}.)
The first term on the right-hand-side of Eq. (\ref{entropy})
is one quarter of the surface area (in $G_n=1$ units); thus the second term
represents a deviation from the one in general relativity. 
This discrepancy gives rise to the negative entropy for the $k=-1$ case, 
which is a characteristic property of any higher-curvature gravitational theories 
because such a deviation traces back its origin to
the terms in the Lagrangian other than the Einstein-Hilbert term
\cite{iyerwald1994}. 
Observe that Eq. (\ref{entropy}) does not reproduce the general
relativistic result in four dimensions, in which the Gauss-Bonnet term
does not alter the dynamics. The correct expression in four dimensions is obtained 
by setting $n=4$ in Eq. (\ref{1stlaw2}) and integrating it.

In Proposition \ref{arealaw}, we have shown the
area law under the null energy condition, 
implying the entropy-increasing law in the general relativistic case.
It deserves to be noted that, since the entropy of a trapping horizon 
$S_{\rm TH}$ is not simply proportional to 
the area in Einstein-Gauss-Bonnet gravity, 
the entropy law--corresponding to the second law of
black-hole mechanics-- is quite nontrivial.

\begin{Prop}
\label{entropylaw}
({\it Entropy law.}) 
{Under the null energy condition}, the entropy of a future outer 
(inner) trapping horizon is non-decreasing
 (non-increasing) along the generator of the 
trapping horizon in both branches.
\end{Prop}
\vspace{5mm}
{\it Proof}. 
From Eq.~(\ref{1stTH}), we obtain
\begin{align}
{\mathscr L}_\xi S_{\rm TH} =
\frac{V_{n-2}^kr_{\rm h}^{n-4}(
r_{\rm h}^2+2{\tilde\alpha}k)}{4 G_n}\theta _- \xi ^u
\label{d-entropy}
\end{align}
along the generator of the trapping horizon.
Repeating the identical procedure given in Proposition \ref{arealaw},
the result follows from Proposition~\ref{th:absenceTH}.
\qed 

\bigskip
It turns out that 
the dynamical entropy $S_{\rm TH}$ increases while
the area of a future outer trapping horizon decreases in the non-GR branch
(trapping horizons occur for only $k=-1$ and $r_{\rm h}^2<2{\tilde\alpha}$). 
This is the exceptional case, appearing only in Einstein-Gauss-Bonnet
gravity. 
When $K^a$ is a Killing vector on $(M^2, g_{ab})$,
we have $\psi ^a=0$, as has been mentioned before, and 
Eq. (\ref{1stTH}) yields the constancy of the entropy
along the trapping horizon.

In closing this section, we summarize the results obtained here in Table~\ref{table:horizons}.
\begin{widetext}
\begin{center}
\begin{table}[h]
\caption{\label{table:horizons} Properties of the future 
trapping horizon under the null energy condition. Each quoted term
denotes that it has the meaning of time evolution only if the trapping
horizon is null, since the area and entropy laws are formulated along the
 generator of the trapping horizon.}
\begin{tabular}{l@{\quad}|@{\quad}c@{\qquad}c@{\quad}|@{\quad}c@{\qquad}c}
\hline \hline
&\multicolumn{2}{c|@{\quad}}{GR branch} &
\multicolumn{2}{c}{non-GR branch} \\
 & future outer & future inner & future outer & future inner \\\hline
signature & non-timelike &  non-spacelike &  non-spacelike & non-timelike \\\hline
trapped side & future & exterior & interior & past \\\hline
area law & ``non-decreasing'' & non-increasing  & non-increasing & ``non-decreasing'' \\\hline
entropy law & ``non-decreasing'' & non-increasing & non-decreasing & ``non-increasing'' \\ \hline
\hline
\end{tabular}
\end{table} 
\end{center}
\end{widetext}

\section{Summary and discussion}
\label{sec:summary}

In this paper, we investigated several aspects of 
dynamical black holes in Einstein-Gauss-Bonnet gravity.  
Properties of trapping horizons were elucidated
for their types and branches. Let us summarize the upshots briefly.
We supposed that the spacetime has symmetries corresponding to 
the isometries of an $(n-2)$-dimensional constant curvature space.
We also assumed the Gauss-Bonnet coupling $\alpha $ to be in the range
$\alpha \ge 0$ and $1+4\tilde \alpha \tilde \Lambda \ge 0$, the
first is motivated by string theory and the second is to
avoid ghosts. 

The quasi-local mass of a trapping horizon 
was shown to obey an inequality summarized in Tables~\ref{table:massgr}--\ref{table:mass-s}. 
In the GR branch with $k=1$ and $\Lambda \le 0$, in particular, 
the quasi-local mass on an untrapped hypersurface with a marginal
surface as an inner boundary has a positive lower bound under the
dominant energy condition, corresponding to  
the value of the mass on the marginal surface.
This isoperimetric inequality is similar to the Penrose inequality 
and establishes the positive mass theorem of a black hole in the case where $k=1$ and $\Lambda \le 0$.

Trapping horizons in the GR branch were shown to 
inherit characteristic properties of those in general relativity. 
If the dominant energy condition and $\Lambda \ge 0$ holds, the topology of outer trapping horizons must have a non-negative curvature.
A future outer trapping horizon is non-timelike under the null energy
condition, embodying the idea that
a black-hole horizon is a one-way membrane. 
Then, the area and entropy laws indicate that a black
hole grows and mimics the second law of thermodynamics, respectively.

In contrast, in the non-GR branch, trapping horizons have some features which
may run counter to our intuition. 
The non-spacelike character of a future outer trapping horizon 
does not display characteristics of a black
hole as a region of no escape. 
Besides that, the area of a future outer trapping horizon 
is non-increasing into the future under the null energy
condition, which does not follow our intuition, either.

To see this more concretely,
let us consider the Hawking evaporation of a black hole, 
in which the null energy condition is violated. 
A black hole in the GR branch continues to lose its mass and
reduce its area. In other words, the signature of a trapping horizon 
becomes non-spacelike and shrinks. 
Whereas in the non-GR branch, a black hole defined by a future outer trapping horizon increases 
its size as it ``evaporates.'' 
A fundamental cause of this
arises from the sign flip in Eq. (\ref{Rvv}) for a radial null vector $k^\mu$,
which makes the non-GR-branch solutions quite eccentric.
But we have not explicitly shown whether this sign change is 
special to radial null vectors or an artifact of our spacetime ansatz (\ref{eq:ansatz}).

We also investigated black-hole dynamics.
In four-dimensional general relativity, 
trapping horizons exhibit laws analogous to black-hole thermodynamics. 
Since their derivation made full use of the 
Einstein equations, it is nontrivial in other theories.
But as shown in \cite{mn2007},
the unified first law strongly suggests the 
first law of a trapping horizon in Einstein-Gauss-Bonnet gravity.
Taking the Kodama vector as an analog of the null generator of 
a Killing horizon,
we defined the surface gravity of a trapping horizon by
following in the footsteps of a Killing horizon. 
The first law of a trapping horizon states that the
energy inflow across the trapping horizon is compensated for
by the entropy gain. 
The resultant dynamical black-hole entropy does not coincide
with one quarter of its area (in $G_n=1$ units), again reproducing Iyer-Wald's result. 
The disagreement with the general relativistic case causes some
annoying but interesting issues. For the $k=-1$ case, for example, the entropy can
be negative. More interestingly, the entropy of a future outer
trapping horizon in the non-GR branch is non-decreasing while 
its area is non-increasing. This is preferable in view of the
second law of black-hole thermodynamics. But the question remains open
whether the generalized second law holds in this system.

It should also be observed that the null trapping horizon 
represent equilibrium configurations. Under the null energy condition, 
its area and entropy are invariant in time, and moreover
energy inflow through the horizon is absent as we have shown below 
Eq. (\ref{expansionderiv}).

We excluded from our consideration the trapping horizons coincident with
the branch points. These trapping horizons occur for $k=-1$, and the areal radius takes
its minimal (maximal) value $r_{\rm h}=\sqrt{2\tilde \alpha }$ in the
GR (non-GR) branch. Since the left-hand-side of Eq. (\ref{Tvv}) becomes identically zero, 
the following propositions could not be established. 
We may truncate these exceptional trapping horizons from our analysis
if the null energy conditions are strictly satisfied, 
because Eq. (\ref{Rvv}) shows that these trapping horizons are singular.
But in other circumstances, we have no definite answer.
The only fact we have at present
is that, from Eq. (\ref{d-entropy}), the entropy remains 
constant along the generator of these trapping
horizons, owing to the vanishing of $\psi_a$ at that point [see
Eq. (\ref{derivk2})]. But we have not understood so far how much 
generality and physical significance these trapping horizons have.

We conclude this paper by commenting on the generalization of the present work. 
When we do not assume the present spacetime symmetries
(\ref{eq:ansatz}),
a more general definition of quasi-local mass is required.
A naive prescription is to generalize the
Hawking mass \cite{Hawking:1968qt}, which should satisfy 
monotonicity and positivity, and represent the
higher-dimensional ADM mass at spatial infinity in the asymptotically flat
case. 
And moreover, this generalization should satisfy the energy balance law
as Eq. (\ref{1stlaw1}). 
We envisage that the extra work terms 
due to the gravitational radiation should be
accompanied in Eq. (\ref{1stTH}), as in the general relativistic case
\cite{Hayward:2004dv,Hayward:2004fz}.
These speculations are challenging but interesting issues 
for future investigations. 
Our analysis should be helpful to those pursuing general properties of trapping
horizons in Einstein-Gauss-Bonnet gravity. 
Extensions into Lovelock gravity
\cite{lovelock} are also an intriguing subject. 
We anticipate that the analysis in the present paper can be extended in Lovelock gravity
and also our quasi-local mass formalism will be of use in analyzing 
the characteristic singularity structure \cite{maeda2006,maeda2006b,nm2006}.
These prospects are left for possible future investigations.


\section*{Acknowledgments}

The authors appreciate a lot of fruitful discussions with Takashi Torii.
MN would like to thank Kei-ichi Maeda for continuous encouragement. 
MN was partially supported by JSPS.
HM was supported by Grant No. 1071125 from FONDECYT (Chile) and the 
Grant-in-Aid for Young Scientists (B), 18740162, from the Scientific 
Research Fund of the Ministry of Education, Culture, Sports, Science and 
Technology (MEXT) of Japan. CECS is funded in part by an institutional 
grant from Millennium Science Initiative, Chile, and the generous 
support to CECS from Empresas CMPC is gratefully acknowledged.


\end{document}